# Mapping the absolute magnetic field and evaluating the quadratic Zeeman effect induced systematic error in an atom interferometer gravimeter

Qing-Qing Hu[1,2,†], Christian Freier[1], Bastian Leykauf[1], Vladimir Schkolnik[1], Jun Yang[2], Markus Krutzik[1,] Achim Peters[1]

[1]*Institut für Physik, Humboldt-Universität zu Berlin, 12489 Berlin, Germany*
[2]*Interdisciplinary Center for Quantum Information, National University of Defense Technology, Changsha, Hunan, 410073, China*

**Abstract**

Precisely evaluating the systematic error induced by the quadratic Zeeman effect is important for developing atom interferometer gravimeters aiming at an accuracy in the μGal regime ($1\mu Gal=10^{-8}\, m/s^2 \approx 10^{-9}\, g$). This paper reports on the experimental investigation of Raman spectroscopy-based magnetic field measurements and the evaluation of the systematic error in the Gravimetric Atom Interferometer (GAIN) due to quadratic Zeeman effect. We discuss Raman duration and frequency step size dependent magnetic field measurement uncertainty, present vector light shift (VLS) and tensor light shift (TLS) induced magnetic field measurement offset, and map the absolute magnetic field inside the interferometer chamber of GAIN with an uncertainty of 0.72 nT and a spatial resolution of 12.8 mm. We evaluate the quadratic Zeeman effect induced gravity measurement error in GAIN as $2.04\,\mu Gal$. The methods shown in this paper are important for precisely mapping the absolute magnetic field in vacuum and reducing the quadratic Zeeman effect induced systematic error in Raman transition-based precision measurements, such as atomic interferometer gravimeters.

## I.     Introduction

Thanks to the development of laser cooling and trapping techniques [1], precision measurements based on cold atom interferometers (AIs) have been demonstrated remarkable prospects, which stretch from atomic gravimeters [2,3], gravity gradiometers [4,5], gyroscopes [6,7] and atomic clocks [8,9], to the measurement of physical constants, such as fine structure constant [10,11], gravitational constant [5,12], and applications in fundamental physics such as quantum tests of the weak equivalence principle [13].To avoid systematic error induced by the first-order Zeeman effect, the atoms are usually prepared in the $m_F = 0$ sublevel before entering the experiment zone. The quadratic Zeeman effect, however, still leads to a non-negligible systematic error [14]. For example, the error caused by the quadratic Zeeman shift is the largest error among all the systematic effect in rubidium fountain clocks [15] and a challenge for developing Raman transition-based atomic gravimeters aiming at an accuracy in the μGal regime [16,17]. Though this error can be alleviated by the Raman wave vector reversing method [4], it cannot be canceled completely due to the spatial non-overlap of the two interference paths [16], especially in the cases of large momentum transfers [18-20] and long pulse intervals [21]. Therefore, mapping the absolute magnetic field intensity in the interference region, and evaluating the corresponding error is a more accurate method for laboratory research and field applications [14,22].

Magnetically-sensitive atom interferometers [23] and double fountain-based simultaneous differential atom interferometers [17] have been proposed to map the gradient of the magnetic field inside a vacuum chamber. However, because the resonance frequencies of the magnetically-sensitive transitions are sensitive to the magnetic field with a scale of 14 Hz/nT [24], a common magnetic field inhomogeneity of 200 nT [23] will cause a frequency detuning of 2.8 kHz, which could change the transition probability of each Raman pulse and result in a non-negligible error. Besides, the absolute magnetic field map is needed in order to precisely evaluate the Zeeman shift induced systematic error. Therefore, one more process of distinguishing the sign of the magnetic field gradient and calculating its spatial integral is needed. Other methods, such as the weak magnetically-sensitive Zeeman splitting [25] and the Bragg interferometers using the three magnetic states simultaneously [26], require experimental conditions such as 30 ms lin ⊥ lin polarized Raman pulse [25] or Bose-Einstein condensate (BEC) atomic sources [26], which are impractical for GAIN and other systems of the same kind [4,27,28].

---
[†] Corresponding author. E-mail: qingqinghu@physik.hu-berlin.de

Raman spectroscopy-based magnetic field mapping was previously reported with a measurement uncertainty of 20 nT [23] and 0.28 nT [29], respectively. In the latter case, in order to achieve this measurement uncertainty and ensure a good spatial resolution simultaneously, the frequency step from shot to shot is as small as 10 Hz, and a 12 ms Raman π pulse is applied when the atoms reaching their apogee. Therefore, the launch velocity and detection time of atomic cloud, as well as the moment of irradiation by Raman pulses need to be adjusted manually for each launch height at which one intends to measure the magnetic field. Besides, taking the parameters of GAIN for example (quantization magnetic field ~5 μT, interferometer chamber length ~68 cm), the 10 Hz frequency step means a time consumption of ~17 days for mapping the magnetic field inside the interferometer chamber with a spatial resolution of 1 cm, and more time is needed for a better spatial resolution. Apparently, this method takes too long a time and might suffer from the problem of magnetic field drift within one measurement. However, this time consumption can be shortened by one order of magnitude if a larger frequency step of 100 Hz is used. Besides, the shorter pulse enables a better spatial resolution and more sampling points in one transition peak (Fig. 3 (a)-(c)) and this might improve the magnetic field measurement uncertainty. Consequently, there has been some interest in investigating the influences of Raman pulse duration and frequency step size on the Raman spectroscopy-based magnetic field measurement. On the other hand, the vector and tensor ac stark light shifts, which influence the measurement accuracy of Raman spectroscopy-based magnetic field measurement, haven't been discussed before.

In this paper, we report on the experimental investigation of Raman spectroscopy-based magnetic field mapping and the evaluation of quadratic Zeeman effect induced systematic error in GAIN. This paper is structured as follows. Section II briefly introduces the measurement principle, experimental setup and procedure. Section III presents (1): the relationship between measurement uncertainty and Raman pulse duration for different frequency step size; (2) the influence of the vector light shift (VLS) and tensor light shift (TLS); (3) the absolute magnetic field map inside the interferometer chamber of GAIN and its time stability; (4) the quadratic Zeeman shift induced systematic error and the uncertainty of this error. Section IV discusses the principle of nulling the ac stark effect in atom interferometer. Section V summarizes our main results and provides an outlook. The appendix shows the calculation of the polarizabilities, the scalar, vector and tensor light shifts in Raman transitions.

## II. Experimental principle and apparatus

### A. Experimental principle

$^{87}$Rb ground state magnetic sublevels and Raman transition configurations are shown in Fig. 1. The $|F,m_F\rangle$ ground state magnetic sublevel will be shifted by $\Delta E = \mu_B g_F m_F B$ when a static magnetic field $B$ exists, where $F$ is atomic total angular momentum, $m_F = 0, \pm 1, \cdots \pm F$ are the projections of total angular momentum on the quantization axis, $\mu_B$ is the Bohr Magneton, the Landé g factor $g_F$ equals 1/2 for the $F = 2$ state and equals -1/2 for the $F = 1$ state. If the magnetic quantization axis is absent, seven transition peaks (dashed black lines in Fig.1) will be observed by sweeping the relative frequency of the two Raman lasers. However, if a quantization magnetic field parallel to the propagating direction of the Raman beams is applied, and the Raman beams are $\sigma^+-\sigma^+$ or $\sigma^--\sigma^-$ polarized, according to the electric dipole transition selection rules, only three transition peaks ($|F=2,m_F\rangle \leftrightarrow |F=1,m_F\rangle$, $m_F = 0, \pm 1$) can be observed (green lines with double arrows in Fig.1) and the magnetic

quantum numbers $m_F$ remains constant before and after Raman transitions. In this case, the resonance frequencies of the $|F=2, m_F\rangle \leftrightarrow |F=1, m_F\rangle$ transition, denoted as $\omega_{m_F, m_F}$ ($m_F = 0, \pm 1$), can be achieved by fitting the Raman spectrum with [30]:

$$P_1 = \frac{\Omega_{eff}^2}{\Omega_{eff}^2 + (\omega - \Delta E/\hbar - \delta_D - \delta_{AC})^2} \sin^2\left(\tau \sqrt{\Omega_{eff}^2 + (\omega - \Delta E/\hbar - \delta_D - \delta_{AC})^2}\right), \quad (1)$$

where $\Omega_{eff} = \Omega_1^* \Omega_2 / 4\Delta$ is the effective Rabi frequency and $\Omega_i = \frac{\Gamma^2}{2 I_s \delta} I_i$ is the Rabi frequency of the $i$th ($i=1, 2$) Raman beam, $\delta_D$ is the Doppler frequency shift, $\delta_{AC} = \Omega_1^{AC} - \Omega_2^{AC}$ is the differential ac Stark light shift, $\Omega_i^{AC} = \Omega_i^2 / 4\Delta$ is the ac Stark light shift of the $i$th Raman beam, $\Delta$ is the single photon detuning, and $\tau$ is the Raman pulse duration. After achieving the three resonance transition frequencies, the magnetic field intensity at the atom-laser interaction position can be inferred from the frequency difference between $\omega_{1,1}$ and $\omega_{0,0}$, denoted as $B_1$, or from the frequency difference between $\omega_{-1,-1}$ and $\omega_{0,0}$, denoted as $B_{-1}$. Both magnetic field results write:

$$\begin{aligned} B_1 &= (\omega_{1,1} - \omega_{0,0})/\gamma_1 \\ B_{-1} &= (\omega_{0,0} - \omega_{-1,-1})/\gamma_1 \end{aligned}, \quad (2)$$

where $\gamma_1 = 2\pi \cdot 14\,\text{Hz/nT}$ is the first order Zeeman coefficient [24].

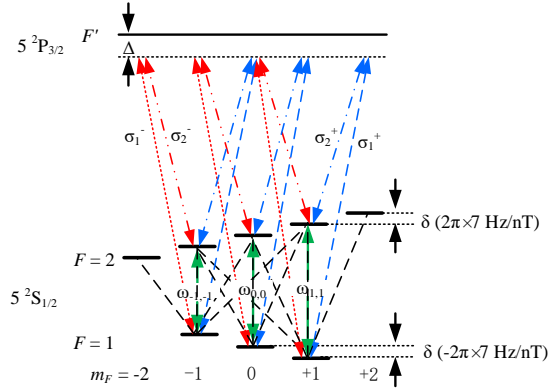

Fig. 1. $^{87}$Rb ground state magnetic sublevels and Raman transition configurations. The green lines with double arrows are possible transitions when a quantization magnetic field parallel to the Raman beams is applied, and the black dashed lines are possible transitions if the quantization axis is absent. $\Delta E$ is the energy level shift induced by first order Zeeman effect, and $\omega_{m_F, m_F}$ ($m_F = 0, \pm 1$) are the resonance frequencies of the $|2, m_F\rangle \to |1, m_F\rangle$ transitions.

## B. Experimental apparatus

The schematic diagram of the experimental setup is shown in Fig. 2 and a detailed description of the whole system can be found in Ref. [31,32]. In this experiment, approximately $10^9$ $^{87}$Rb atoms are first cooled and trapped in the magnetically shielded Magneto-Optical Trap (MOT) chamber within 0.6 s, and then launched vertically by moving molasses technique, achieving a temperature of ~2 μK and an initial velocity of ~4.4 m/s in 3 ms. The repumping laser, whose frequency is near resonance with $|F=1\rangle \to |F'=2\rangle$ transition frequency, is switched off 1 ms later than the cooling laser to ensure all atoms are in the $F=2$ ground state. The collimated Raman π pulse with an $e^{-2}$ diameter of 29.5 mm are irradiated from the top vacuum window while the atoms are moving in the magnetically shielded interferometer chamber. As shown in Fig.2, the magnetic shield consists of three equally spaced concentric cylinders made from 0.75 mm thick sheets of high-permeability nickel-iron-molybdenum alloy. The innermost cylinder has a length of 68 cm and a diameter of 8 cm while the outermost is 72 cm and 13 cm with the middle and outer cylinder caped to the open diameter of the inner cylinder. This shield provides an attenuation factor of roughly 1000 for background magnetic field. A solenoid, driven by a precision laser diode current driver whose root of mean square (RMS) current noise is ~1 µA, is wound precisely inside the inner mu-metal shield to create a highly homogeneous quantization magnetic field in vertical direction. When the atoms fall down through the detection chamber, a normalized fluorescence detection process is applied. Specifically, an 18 mm $e^{-2}$ diameter detection pulse whose frequency is near resonance with the $|F=2\rangle \to |F'=3\rangle$ transition frequency is applied for 320 µs in horizontal plane and a photomultiplier tube is applied in orthogonal direction to detect the atom population $N_2$ in $F=2$ state. Afterwards, a 60 µs repumping beam and a second 320 µs detection pulse are applied orderly to measure the total atom number $N_1 + N_2$. The Raman transition probability can be inferred by $P_1 = 1 - N_2/(N_1 + N_2)$. The whole launch-detection experimental cycle time is ~1.5 s.

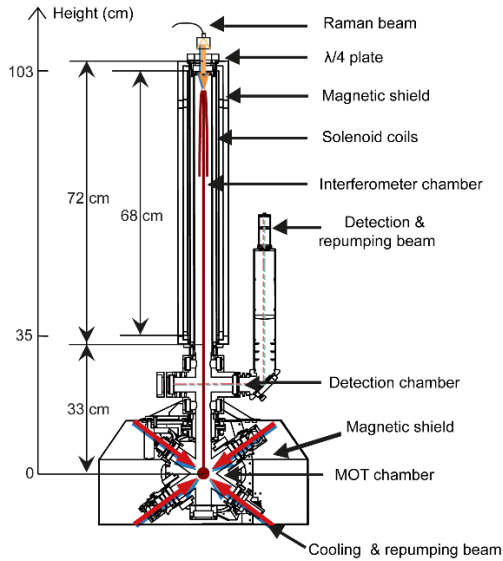

Fig. 2. Schematic diagram of the experimental setup.

### III. Experimental results

#### A. The Raman pulse duration and frequency step dependent magnetic field measurement uncertainty

To investigate the influences of Raman pulse duration $\tau$ and frequency step size $\Delta f$ on the Raman spectroscopy-based magnetic field measurement, we first measured the magnetic field at one fixed position (the same timing for Raman pulse) with different Raman pulse duration and frequency step size. The obtained Raman spectra is shown in Fig. 3, in which Fig. 3(a)-(c) show the influence of frequency step size for the same Raman pulse duration, Fig. 3(a) and 3(d) (Fig. 3(b) and 3(e)) show the influence of Raman pulse duration for the same frequency step. The three resonance frequencies $\omega_{m_F}$ ($m_F = 0, \pm 1$) of each Raman spectrum are obtained by fitting it with a combined Raman transition function of Eq. (1) (red lines in Fig. 3). The RMS fitting uncertainty of the three resonance frequencies divided by $\gamma_1$ is defined as the magnetic field measurement uncertainty. This fitting uncertainty is also confirmed by the Bootstrap method[33].

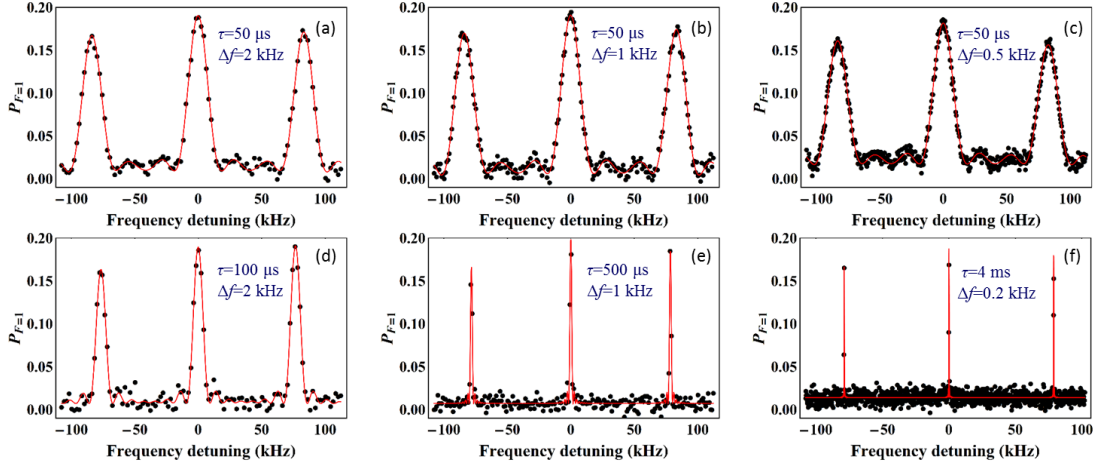

Fig. 3. Raman spectra obtained at one fixed height with different Raman duration $\tau$ and frequency step $\Delta f$. Black points are measured data and red lines are fitting results with a combined function of Eq. (1).

The dependencies of measurement uncertainty on Raman pulse duration and frequency step size are shown in Fig. 4. When the frequency domain sampling theorem is fulfilled ($\Delta f \leq 1/(2\tau)$), the measurement uncertainty improves with decreasing frequency step size $\Delta f$ approximatively as $U_B \propto \sqrt{\Delta f}$, which matches the relationship of $U_B \propto 1/\sqrt{N}$, considering the number of sampling points $N \propto \Delta f^{-1}$. As a result of fewer sampling points in the transition peaks when increasing the Raman pulse duration with the same frequency step size, the steepness of the measurement uncertainty is smaller than $1/\tau$ (black dotted line with crosses in Fig. 4). Alternatively, by choosing a suitable scanning region, i.e., taking data only within the position of the Raman peaks, one could improve these results, getting closer to the expected $1/\tau$ behavior. It's worth mentioning that when $\Delta f = 25\,\text{Hz}$, $\tau = 4\,\text{ms}$, the achieved magnetic field measurement uncertainty is 0.157 nT, which is better than the best Raman spectroscopy-based magnetic field measurement result of 0.28 nT we found in Ref. [29]. An even lower measurement uncertainty can be achieved by using a smaller $\Delta f$ together with a longer $\tau$. However, a longer $\tau$ corresponds to a lower spatial resolution (assuming a fixed launch velocity), and a smaller $\Delta f$ increases the time needed for a complete scan of the spectrum and thus the time needed to map the magnetic field inside the whole interferometer chamber. Therefore, a compromise should be made between the measurement uncertainty, the spatial resolution and the time consumption on the choice of the experimental parameters for mapping the magnetic field of the whole interferometer chamber, especially when the measurement time is limited, such as in field applications.

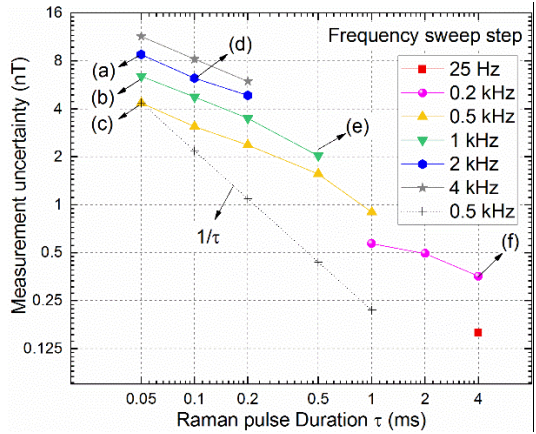

Fig. 4. Magnetic field measurement uncertainty as a function of Raman pulse duration $\tau$ for different frequency step $\Delta f$. When the frequency domain sampling theorem is fulfilled ($\Delta f \leq 1/(2\tau)$), the measurement uncertainty improves with decreasing $\Delta f$ approximatively as $U_B \propto \sqrt{\Delta f}$ while improves with increasing $\tau$ with a smaller steepness than $1/\tau$. The measured data are represented by the points, and the lines are simply to guide the eye. The data marked with (a)-(f) represent the achieved measurement uncertainties corresponding to the spectrums shown in Fig. 3(a)-3(f).

**B. The vector and tensor light shift induced magnetic field measurement offset**

From Eq. (1) and (2) we can see that the magnetic field measurement accuracy is mainly affected by the Doppler shift and ac stark shift. In copropagating Raman configuration, the Doppler effect is very small (223.7 mHz during the 1 ms Raman pulse) compared to the 1st order Zeeman frequency shift (~70 kHz), and is canceled further by calculating the difference of the adjacent transition peaks because they are identical for the three transition peaks (Fig. 3). The ac Stark energy shift on the hyperfine-structure state $|Fm_F\rangle$ induced by laser field $E(t) = \varepsilon_L \hat{\zeta} e^{-i(\omega_L t + \hat{k} \cdot \hat{r})} + c.c.$ is [9,34-36]

$$\delta E^{AC}_{Fm_F} = -\left(\frac{\varepsilon_L}{2}\right)^2 \left[\alpha_F^S(\omega_L) + (\hat{k} \cdot \hat{B})\mathbb{A}\frac{m_F}{2F}\alpha_F^V(\omega_L) + \left(3|\hat{\zeta} \cdot \hat{B}|^2 - 1\right)\frac{3m_F^2 - F(F+1)}{2F(2F-1)}\alpha_F^T(\omega_L)\right] + O^{(4)}, \quad (3)$$

where the superscripts $S$, $V$, and $T$ distinguish the scalar, vector, and tensor parts of the polarizability $\alpha_F(\omega)$. $\hat{k}$ and $\hat{B}$ are unit vectors along the laser wave vector and quantization magnetic field, respectively. $\mathbb{A}$ is the degree of circular polarization of the laser [37]: $\mathbb{A} = \pm 1$ for $\sigma^\pm$ laser and $\mathbb{A} = 0$ for a linearly polarized laser (the vector light shift drops out in this case). $\hat{\zeta}$ is the complex polarization vector of the laser and may be expressed as $\hat{\zeta} = e^{i\gamma}\left(\cos\varphi\hat{\zeta}_{maj} + i\sin\varphi\hat{\zeta}_{min}\right)$, where $\gamma$ is a real, the real unit vectors $\hat{\zeta}_{maj}$ and $\hat{\zeta}_{min}$ ($\hat{\zeta}_{maj} \times \hat{\zeta}_{min} = \hat{k}$) align with the semi-major axis and semi-minor axis of the ellipse which swept out by the electric field vector of laser in one period, respectively, and $\varphi$, defined by $\tan\varphi = |\hat{\zeta}_{min}|/|\hat{\zeta}_{maj}|$, is directly related to the degree of circular polarization $\mathbb{A}$. More information can be find in Figure 3.3 of Ref. [37]. $O^{(4)}$ represents the higher order terms.

From Eq. (2) and (3), it's clear that the scalar light shifts (SLS) are identical for all the three magnetic states ($m_F = 0, \pm 1$) thus is canceled out by calculating the frequency difference of the adjacent transition peaks (Fig. 3). When the Raman laser is circular polarized ($\mathbb{A} = \pm 1$), the vector light shift (VLS) will be opposite for the $m_F = +1$ and $m_F = -1$ (magnetically-sensitive) states [25] but equals 0 for the $m_F = 0$ (magnetically-insensitive) state, thus will manifest as a "fictitious" magnetic field $B_{vls}$. Assuming the intensity ratio of the two Raman lasers is $I_1/I_2 = q$, calculating the frequency shift in Raman transition caused by VLS using Eq. (7.471) of Ref. [35], and taking the Zeeman splitting $\delta E^B_{Fm_F} = \mu_B g_F m_F B$ into consideration, the VLS induced "fictitious" magnetic field in $\Delta m_F = 0$ Raman transition can be written as:

$$B_{vls} = \frac{\delta E^V_{m_F}}{\mu_B m_F} = -\mathbb{A}\eta\varepsilon_{L_2}^2, \quad (4)$$

where $\eta$ is a constant determined by the frequencies and intensity ratio of the two Raman lasers (see Eq. (A11) in Appendix). Eq. (4) shows that a polarization dependent "fictitious" magnetic field $B_{vls}$ is created in the circular polarized Raman configuration ($A = \pm 1$), and the amplitudes of $B_{vls}$ is inversely proportional to the Raman duration $\tau$ as $\tau = \pi/\Omega_{eff} \propto \varepsilon_{L_2}^{-2}$. The measured magnetic field intensities $B_{m_F}^{\sigma}$ are shown as points in Fig. 5, in which the superscript $\sigma$ ($\sigma = \sigma^+, \sigma^-$) represents the polarization of Raman laser and the subscript $m_F$ represents the magnetic field inferred from the resonance frequency of the $m_F$ magnetic state, respectively. The fitting results show the absolute magnetic field is 5651.7 nT and the VLS induced offset is 26.8 nT when the circular polarized ($A = \pm 1$) Raman pulse duration is 1 ms.

From Eq. (2) and (3), we can see that the tensor light shifts (TLS) are identical for $m_F = +1$ and $m_F = -1$ states but different for $m_F = 0$ state, thus will lead to a difference between $B_1$ and $B_{-1}$. This difference is proportional to the intensity of Raman laser but equal for both left- and right-handed circularly polarized Raman laser beams in our experiments. The tensor polarizability $\alpha_F^T(\omega)$ calculated from Eq. (7.471) of Ref. [35] is about one order of magnitude smaller than the vector polarizability and tends to zero when the laser is far-detuned (see Eq. (12)-(13)), thus the effect of TLS (the difference between magnetic fields inferred from different magnetic states, i.e., $B_1^{\sigma+}$ and $B_{-1}^{\sigma+}$) is one order of magnitude smaller than the effect of VLS (the difference between magnetic fields measured with different Raman laser polarization, i.e., $B_1^{\sigma+}$ and $B_1^{\sigma-}$), as shown in Fig.5. By calculating the average value of the measured results for both $\sigma^-\sigma^-$ and $\sigma^+\sigma^+$ polarization configuration (orange stars), the polarization dependent VLS can be canceled and the effect of TLS can be extracted. The fitting result (orange line) shows that the TLS induced magnetic field offsets are 2.2 nT when the Raman pulse duration is 1 ms and 44.3 nT when the Raman pulse duration is 50 μs, respectively.

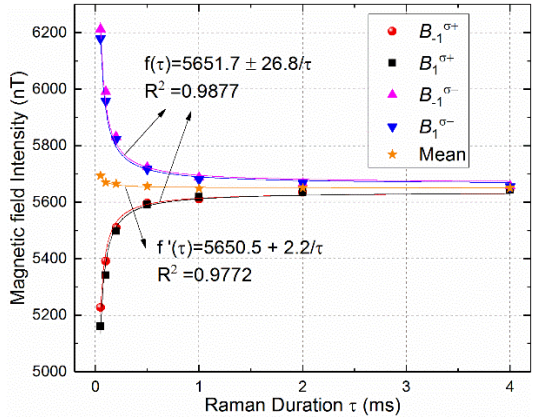

Fig. 5. The influence of vector and tensor light shift on measured magnetic field intensity. The measured results $B_{m_F}^{\sigma}$ are shown as points, in which the superscript $\sigma$ ($\sigma = \sigma^+, \sigma^-$) represents the polarization of Raman laser, and the subscript $m_F$ represents the result inferred from the resonance frequency of the $|F=2, m_F\rangle \leftrightarrow |F=1, m_F\rangle$ transition peak, respectively. The amplitude of the "fictitious" magnetic field induced by VLS is inversely proportional to the Raman duration as $B_{vls} \propto 1/\tau$. The effect of TLS is one order of magnitude smaller than the effect of VLS.

### C. Magnetic field map inside the interferometer chamber and its stability

Here we choose the parameters of 1 ms pulse length and 400 Hz frequency step size to map the magnetic field inside the 68-cm-high interferometer chamber of GAIN. These parameters correspond to a time consumption of ~12.5 minutes (200 kHz frequency sweep range) and a measurement uncertainty of 0.72 nT (a sensitivity of 19.7 $nT/\sqrt{Hz}$) for each measurement. In order to map the magnetic field inside the whole interferometer chamber, one has to irradiate the atoms with Raman pulse at different times on the atom's trajectory. We choose a time delay of 4 ms for Raman pulse irradiation between measurements, resulting in 69 measurement heights (~14 hours) in total. The whole magnetic field mapping process is implemented automatically by making a 2-Dimension scan (Raman laser frequency scan and Raman laser irradiation time scan). In order to obtain the absolute magnetic field intensity and identify the source of the field inhomogeneity, we implemented the above mapping process four times with nominal (13 mA) and half (6.5 mA) solenoid currents for $\sigma^-\sigma^-$ and $\sigma^+\sigma^+$ polarization configurations, respectively.

The achieved absolute magnetic field maps of the interferometer chamber of GAIN for nominal and half solenoid currents are shown in Fig. 6 (a) and (b), in which the height is referred to the center of the MOT chamber (Fig.2). The spatial resolution of the magnetic field maps is determined by three factors: atom's flight distance during the 1 ms Raman pulse, atom's flight distance during the 4 ms Raman pulse delay, as well as atomic cloud's diameter at the time of Raman pulse. We here take atom's largest flight distance of 12.8 mm during the 4 ms Raman pulse delay time as a lower limit of the spatial resolution.

Considering the magnetic field in vertical direction consists of solenoid magnetic field ($B_{sn}$) and background magnetic field ($B_{bg}$) (the horizontal component of $B_{bg}$ is very small and omitted here), the measured magnetic fields for nominal ($B_N$) and half ($B_H$) solenoid current can be written as

$$B_N = B_{sn} + B_{bg}, \tag{5}$$

$$B_H = 0.5 \cdot B_{sn} + B_{bg}. \tag{6}$$

where the offset of the current driver, which may result in some offset in quantization magnetic field and need to be analyzed later in detail, is omitted here. Therefore, $B_{sn}$ and $B_{bg}$ can be inferred from

$$B_{sn} = 2(B_N - B_H), \tag{7}$$

$$B_{bg} = 2B_H - B_N. \tag{8}$$

The inferred solenoid magnetic field $B_{sn}$ is very homogeneous with a standard deviation (SD) of 6.24 nT (Fig. 6(c)), while the inferred background magnetic field $B_{bg}$ has a similar fluctuation of about 100 nT (Fig. 6(d)) as the measured magnetic fields $B_N$ and $B_H$ (Fig. 6(a) and 6(b)). Therefore, the magnetic field inhomogeneity in the interferometer zone is attributed to the residual background magnetic field.

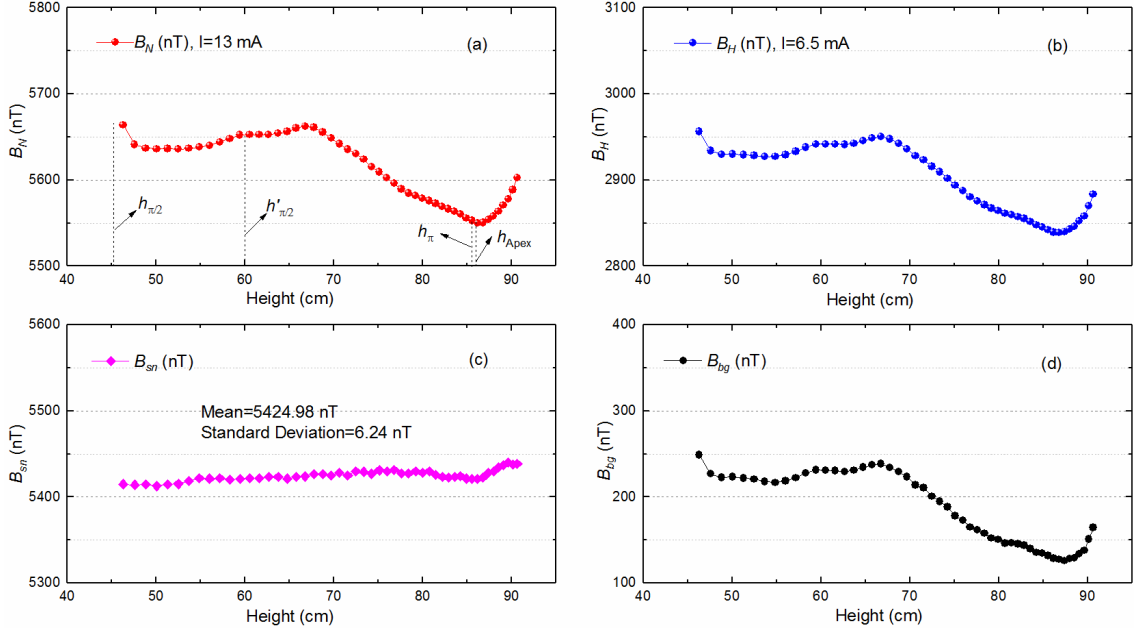

Fig. 6. Measured magnetic field map for the (a) nominal (13 mA) and (b) half (6.5 mA) solenoid current; Inferred field map for the (c) solenoid and (d) background magnetic field. The data are represented by the points, and the lines are simply to guide the eye. Relevant heights for our standard atom interferometer configuration are indicated in (a).

In order to evaluate the stability of the background magnetic field and current driver, we measure continuously at one fixed position with nominal solenoid current for 12.5 hours (the experimental parameters are same as the mapping process, namely 1 ms Raman pulse duration and 400 Hz frequency step). The Allan Deviation of the measured magnetic field decreases to ~0.4 nT after $2\times10^3$ seconds and decreases again for longer averaging time after a small increase, indicating the stability of the current driver and background magnetic field, as well as the measurement precision of this method.

### D. Quadratic Zeeman effect induced systematic error in GAIN

With the knowledge of the magnetic field intensity the atoms experienced during flight, denoted as $B(t)$, the quadratic Zeeman effect induced phase shift $\Delta\phi_{zeeman}$ and gravity error $\Delta g$ in atom interferometer gravimeters can be inferred from

$$\Delta\phi_{zeeman} = \gamma_2 \int_{-T}^{T} g_s(t) B^2(t) dt, \qquad (9)$$

$$\Delta g = \frac{\Delta\phi_{zeeman}}{k_{eff} T^2}, \qquad (10)$$

where $\gamma_2 = 2\pi \times 0.0575\,\text{Hz/nT}^2$ is the quadratic Zeeman coefficient, $g_s(t)$ is the sensitivity function [38] of the Mach-Zehnder (M-Z) atom interferometer, and $T$ is the time interval between pulses.

As shown in Eq. (9), in order to decrease the influence of the quadratic Zeeman effect, GAIN is implemented in fountain configuration in which situation the flight trajectory of the atoms during the first half and second half of the interferometer path of the Mach-Zehnder (M-Z) atom interferometer is almost symmetric (see Fig. 6(a)), with a small discrepancy from the fountain apex to ensure the atoms have a non-negligible velocity for Doppler-sensitive Raman transition. With typical experimental parameters of $v_{launch} = 4.1 \, \text{m/s}$, $t_{\pi/2} = 0.73 \, \text{s}$, $T = 0.26 \, \text{s}$, ($h_{\pi/2} \approx 45.1 \, \text{cm}$, $h_{\pi} \approx 85.7 \, \text{cm}$, $h'_{\pi/2} \approx 60.0 \, \text{cm}$, $h_{Apex} \approx 86.1 \, \text{cm}$, see Fig. 6(a)), the gravity offset $\Delta g$ inferred by the interpolation integral of the magnetic field map for the nominal solenoid current is $\Delta g = 2.04 \, \mu\text{Gal}$. The uncertainty of the gravity offset due to the uncertainty of the magnetic field, inferred from $U_{\Delta g} \approx 2\Delta g \, U_B / \bar{B}$, is $0.52 \, \text{nGal}$. Therefore, the experimental parameters of 1 ms pulse duration and 400 Hz frequency step size are sufficient for subtracting the quadratic Zeeman effect related systematic error to an uncertainty of nGal level.

For comparison, if this atom interferometer would be implemented in a free fall configuration (i.e., releasing the atoms from a MOT at the top of the instrument), then the geometric symmetry of atom's moving trajectory will be lost and the interval time between the Raman pulses will be decreased to $T' = 0.14 \, \text{s}$, the gravity error caused by the magnetic field inhomogeneity will be enlarged to $\Delta g' = 12.47 \, \mu\text{Gal}$ with an enlarged uncertainty of $U_{\Delta g'} = 3.20 \, \text{nGal}$.

## IV. Discussion

For atom interferometer gravimeter, the intensity ratio $q$ of the two Raman lasers is usually set to a particular value in order to cancel the influence of the light shift [39]. Taking the experimental parameters of GAIN as example, in which $\omega_1 = \omega_{1'2} - 700 \, \text{MHz}$, $\omega_2 = \omega_1 + 6.834 \, \text{GHz}$, the differential energy shifts induced by the scalar, vector and tensor light shifts of the $|F=1, m_F\rangle \leftrightarrow |F=2, m_F\rangle$ Raman transitions (see Eq. (A8) - (A10) in Appendix) can be simplified to:

$$\delta E^S = -\left(\frac{\varepsilon_{L2}}{2}\right)^2 h\{8.327q - 14.814\}, \qquad (11)$$

$$\delta E^V_{m_F} = -\left(\frac{\varepsilon_{L2}}{2}\right)^2 \mathbb{A} m_F h\{1.069q + 2.123\}, \qquad (12)$$

$$\delta E^T_{m_F} = -\left(\frac{\varepsilon_{L2}}{2}\right)^2 \left(3|\hat{\zeta} \cdot \hat{B}|^2 - 1\right) h\{-0.168 - 0.207q + m_F^2(0.245 + 0.105q)\}. \qquad (13)$$

where the propagation direction of the Raman laser is assumed parallel to the quantization magnetic field, namely $\hat{k} \cdot \hat{B} = 1$ in Eq. (A9).

Usually, the polarization (propagation) direction of the Raman laser is perpendicular (parallel) to the magnetic quantization axis, corresponding to $|\hat{\zeta} \cdot \hat{B}|^2 = 0$ in Eq. (13), thus the total differential light shift of the $|F=1, m_F\rangle \leftrightarrow |F=2, m_F\rangle$ Raman transition can be inferred from Eq.(11)-(13) as:

$$\delta E^{Total}_{m_F} = -\left(\frac{\varepsilon_{L2}}{2}\right)^2 h\{8.534q - 14.646 + m_F \mathbb{A}(2.123 + 1.069) - m_F^2(0.245 + 0.105q)\}. \qquad (14)$$

Furthermore, the Raman lasers of the atom interferometers are usually in $\text{lin} \perp \text{lin}$ polarization configurations [39], correspond to $\mathbb{A} = 0$, and the atoms are prepared in the magnetically-insensitive state, correspond to $m_F = 0$. According to Eq. (11)-(14), the vector light shift equals zero, and the total (including the scalar and tensor) light shift can be canceled by setting the intensity ratio of the two Raman lasers to $q = 1.718$. It's worth mentioning that this result of $q = 1.718$ is more closer to the experimental result of $q = 1.72$ shown in Fig. 5.7 of Ref. [39] than the result of $q = 1.779$ obtained when neglecting the tensor polarizability.

Eq. (11) – (14) also show that in order to cancel the influence of light shift, the Raman laser intensity ratio $q$ for atomic interferometers with atoms in magnetically-sensitive states ($m_F = \pm 1$) is different from the intensity ratio $q'$ for atomic interferometers with atoms in magnetically-insensitive state ($m_F = 0$) considering the contribution of the tensor light shift. Furthermore, for atom interferometer with magnetically-sensitive atom states ($m_F = \pm 1$), the contribution of the vector light shift should be considered as well if the Raman laser is circular polarized ($A = \pm 1$).

## V. Conclusion and outlook

In this paper, we reported on the experimental investigation of Raman spectroscopy-based magnetic field mapping method and the evaluation of quadratic Zeeman effect induced systematic offset in the Gravimetric Atom Interferometer (GAIN). We show both Raman pulse duration and frequency step size dependent measurement uncertainty, investigated the influence of vector light shift (VLS) and tensor light shift (TLS), and presented a method to extract the absolute magnetic field intensity and the TLS. We mapped the absolute magnetic field inside the interferometer chamber of GAIN automatically with 1 ms Raman π pulse and 400 Hz frequency step, achieving a magnetic field measurement uncertainty of 0.72 nT and a spatial resolution of lower than 12.8 mm. We attributed the magnetic field inhomogeneity of ~100 nT to the residual background magnetic field which can be decreased further by improving the magnetic shield. The quadratic Zeeman effect induced gravity measurement offset in GAIN is evaluated as 2.04 µGal, in which the offset of the current driver and other error sources still need to be analyzed later in detail. The methods shown in this paper can be used for precisely mapping the absolute magnetic field in vacuum and reducing the systematic error budget in Raman transition-based precision measurements, such as atomic interferometer gravimeters.

### Acknowledgments

This material is based on work funded by the European Commission (FINAQS, Contr. No. 012986-2 NEST), by ESA (SAI, Contr. No. 20578/07/NL/VJ) and by ESF/DFG (EuroQUASAR-IQS, DFG grant PE 904/2-1 and PE 904/4-1). Qingqing Hu would like to thank the support of the National Natural Science Foundation of China under Grant No. 51275523, Specialized Research Fund for the Doctoral Program of Higher Education of China under Grant No. 20134307110009, and the Graduate Innovative Research Fund of Hunan Province under Grant No. CX2014A002.

### Appendix: Calculation of the polarizabilities and light shifts in Raman transition

From Eq.(7.471) of Ref. [35], the scalar, vector, and tensor polarizabilities of atom in ground state $|F\rangle$ are:

$$\alpha_F^S(\omega) = \sum_{F'} \frac{2\omega_{F'F} |\langle F\|d\|F'\rangle|^2}{3\hbar(\omega_{F'F}^2 - \omega^2)}, \tag{A1}$$

$$\alpha_F^V(\omega) = \sum_{F'} (-1)^{F+F'+1} \sqrt{\frac{6F(2F+1)}{F+1}} \begin{Bmatrix} 1 & 1 & 1 \\ F & F & F' \end{Bmatrix} \frac{\omega_{F'F} |\langle F\|d\|F'\rangle|^2}{\hbar(\omega_{F'F}^2 - \omega^2)}, \tag{A2}$$

$$\alpha_F^T(\omega) = \sum_{F'} (-1)^{F+F'} \sqrt{\frac{40F(2F+1)(2F-1)}{3(F+1)(2F+3)}} \begin{Bmatrix} 1 & 1 & 2 \\ F & F & F' \end{Bmatrix} \frac{\omega_{F'F} |\langle F\|d\|F'\rangle|^2}{\hbar(\omega_{F'F}^2 - \omega^2)}, \tag{A3}$$

where $\omega_{F'F}$ represents the resonant laser frequency of ground state $F$ to excited state $F'$, $\omega$ is the frequency of laser, $|\langle F\|d\|F'\rangle|$ is the reduced matrix element [24], $\begin{Bmatrix} 1 & 1 & 1 \\ F & F & F' \end{Bmatrix}$ and $\begin{Bmatrix} 1 & 1 & 2 \\ F & F & F' \end{Bmatrix}$ are the Wigner 6-j symbols [35].

By calculating the Wigner 6-j symbols, and taking the reduced matrix elements and the resonant laser frequencies $\omega_{F'F}$ of $F \to F'$ dipole transitions from Ref. [24], the scalar, vector, and tensor polarizabilities of the $^{87}$Rb atom in $F=1$ and $F=2$ ground states can be simplified to:

$$\alpha_1^S(\omega) = \left[\frac{\omega_{01}}{9(\omega_{01}^2 - \omega^2)} + \frac{5\omega_{11}}{18(\omega_{11}^2 - \omega^2)} + \frac{5\omega_{21}}{18(\omega_{21}^2 - \omega^2)}\right]\frac{d_2^2}{\hbar}$$

$$\alpha_2^S(\omega) = \left[\frac{\omega_{12}}{30(\omega_{12}^2 - \omega^2)} + \frac{\omega_{22}}{6(\omega_{22}^2 - \omega^2)} + \frac{7\omega_{32}}{15(\omega_{32}^2 - \omega^2)}\right]\frac{d_2^2}{\hbar}, \quad (A4)$$

$$\alpha_1^V(\omega) = \left[-\frac{\omega_{01}}{6(\omega_{01}^2 - \omega^2)} - \frac{5\omega_{11}}{24(\omega_{11}^2 - \omega^2)} + \frac{5\omega_{21}}{24(\omega_{21}^2 - \omega^2)}\right]\frac{d_2^2}{\hbar}$$

$$\alpha_2^V(\omega) = \left[-\frac{\omega_{12}}{20(\omega_{12}^2 - \omega^2)} - \frac{\omega_{22}}{12(\omega_{22}^2 - \omega^2)} + \frac{7\omega_{32}}{15(\omega_{32}^2 - \omega^2)}\right]\frac{d_2^2}{\hbar}, \quad (A5)$$

$$\alpha_1^T(\omega) = \left[-\frac{\omega_{01}}{9(\omega_{01}^2 - \omega^2)} + \frac{5\omega_{11}}{36(\omega_{11}^2 - \omega^2)} - \frac{\omega_{21}}{36(\omega_{21}^2 - \omega^2)}\right]\frac{d_2^2}{\hbar}$$

$$\alpha_2^T(\omega) = \left[-\frac{\omega_{12}}{30(\omega_{12}^2 - \omega^2)} + \frac{\omega_{22}}{6(\omega_{22}^2 - \omega^2)} - \frac{2\omega_{32}}{15(\omega_{32}^2 - \omega^2)}\right]\frac{d_2^2}{\hbar}, \quad (A6)$$

where $d_2 = |\langle J=1/2\|er\|J'=3/2\rangle|$ is the reduced D2 transition dipole matrix element of $^{87}$Rb.

In the case of two-photon Raman transition, there are two laser fields, $E_1(t)$ and $E_2(t)$. According to the ac Stark energy level shift of hyperfine state $|Fm_F\rangle$ induced by laser field $E(t) = \varepsilon_L \hat{\zeta} e^{-i(\omega_L t + \hat{k}\cdot\hat{r})} + c.c.$, Eq.(3) in the text,

$$\delta E_{Fm_F}^{AC} = -\left(\frac{\varepsilon_L}{2}\right)^2 \left[\alpha_F^S(\omega_L) + (\hat{k}\cdot\hat{B})\mathbb{A}\frac{m_F}{2F}\alpha_F^V(\omega_L) + \left(3|\hat{\zeta}\cdot\hat{B}|^2 - 1\right)\frac{3m_F^2 - F(F+1)}{2F(2F-1)}\alpha_F^T(\omega_L)\right] + O^{(4)}, \quad (A7)$$

the differential energy level shifts on Raman transition caused by scalar, vector, and tensor light shifts can be inferred by first summing the light shifts on $F=1$ state and $F=2$ state induced by the two Raman lasers $E_1(t)$ and $E_2(t)$ with frequencies of $\omega_1$ and $\omega_2$ (define intensity ratio $q = I_1/I_2$) respectively, and then calculating the differential light shifts between $F=1$ state and $F=2$ state, which are written as:

$$\delta E^S = -\left(\frac{\varepsilon_{L2}}{2}\right)^2 \left\{\left[q\alpha_2^S(\omega_1) + \alpha_2^S(\omega_2)\right] - \left[q\alpha_1^S(\omega_1) + \alpha_1^S(\omega_2)\right]\right\}, \quad (A8)$$

$$\delta E_{m_F}^V = -\left(\frac{\varepsilon_{L2}}{2}\right)^2 (\hat{k}\cdot\hat{B})\mathbb{A}m_F \left\{\frac{1}{4}\left[q\alpha_2^V(\omega_1) + \alpha_2^V(\omega_2)\right] - \frac{1}{2}\left[q\alpha_1^V(\omega_1) + \alpha_1^V(\omega_2)\right]\right\}, \quad (A9)$$

$$\delta E_{m_F}^T = -\left(\frac{\varepsilon_{L2}}{2}\right)^2 \left(3\left|\hat{\zeta}\cdot\hat{B}\right|^2 - 1\right)\left\{\frac{3m_F^2-6}{12}\left[q\alpha_2^T(\omega_1)+\alpha_2^T(\omega_2)\right] - \frac{3m_F^2-2}{2}\left[q\alpha_1^T(\omega_1)+\alpha_1^T(\omega_2)\right]\right\}, \quad (A10)$$

where $\alpha_F^l(\omega_j)$, calculated from Eq. (A4) - (A6), represent the scalar $(l = S)$, vector $(l = V)$, and tensor polarizabilities $(l = T)$ of $^{87}$Rb atom's $F = 1$ and $F = 2$ ground states induced by Raman lasers $E_1(t)$ and $E_2(t)$ with frequencies of $\omega_1$ and $\omega_2$. In our experiment, $\omega_1 = \omega_{1'2} - 700\,\text{MHz}$, $\omega_2 = \omega_1 + 6.834\,\text{GHz}$.

Considering the Zeeman splitting $\delta E_{Fm_F}^B = \mu_B g_F m_F B$, where $\mu_B$ is the Bohr Magneton, the Landé g factor $g_F$ equals 1/2 for the $F = 2$ state and equals -1/2 for the $F = 1$ state, the VLS induced "fictitious" magnetic field in $\Delta m_F = 0$ Raman transition can be written as:

$$B_{vls} = \frac{\delta E_{m_F}^V}{\mu_B m_F} = -\mathrm{A}\frac{\left\{\left[q\alpha_2^V(\omega_1)+\alpha_2^V(\omega_2)\right]-2\left[q\alpha_1^V(\omega_1)+\alpha_1^V(\omega_2)\right]\right\}}{16\mu_B}\varepsilon_{L_2}^2 = -\mathrm{A}\eta\varepsilon_{L_2}^2, \quad (A11)$$

where $\eta = \left\{\left[q\alpha_2^V(\omega_1)+\alpha_2^V(\omega_2)\right]-2\left[q\alpha_1^V(\omega_1)+\alpha_1^V(\omega_2)\right]\right\}/16\mu_B$ is a constant determined by the frequencies and intensity ratio of the two Raman lasers.